\documentclass[twocolumn,showpacs,prl]{revtex4}

\usepackage{epsfig,latexsym,graphicx,amsfonts,bm,pifont}
\newcommand{\beq}{ \begin{equation} }
\newcommand{\eeq}{ \end{equation} }
\newcommand{\beqs}{ \begin{eqnarray} }
\newcommand{\eeqs}{ \end{eqnarray} }

\newcommand{\f}[2]{\frac{#1}{#2}}

\newcommand{\sub}[1]{_{\! _{#1}}}

\begin{document}

\title{Dynamics of a deformable body in a fast flowing soap film}

\author{Sunghwan Jung$^1$}
\author{Kathleen Mareck$^1$}
\author{Michael Shelley$^1$}
\author{Jun Zhang$^{2,1}$}
\affiliation{$^1$Applied Mathematics
Laboratory, Courant Institute of Mathematical Sciences, New York
University,
251 Mercer Street, New York, New York 10012, USA \\
$^2$Department of Physics, New York University, 4 Washington
Place, New York, New York 10003, USA}

\date{\today}

\begin{abstract}
We study the behavior of an elastic loop embedded in a flowing
soap film.  This deformable loop is wetted into the film and is
held fixed at a single point against the oncoming flow. We
interpret this system as a two-dimensional flexible body
interacting in a two-dimensional flow. This coupled
fluid-structure system shows bistability, with both stationary and
oscillatory states. In its stationary state, the loop remains
essentially motionless and its wake is a von K\'arm\'an vortex
street. In its oscillatory state, the loop sheds two vortex
dipoles, or more complicated vortical structures, within each
oscillation period. We find that the oscillation frequency of the
loop is linearly proportional to the flow velocity, and that the
measured Strouhal numbers can be separated based on wake
structure.
\end{abstract}

 % 47.32ck: Vortex Streets, 47.50Gj:Instabilities
% 47.20-k:Flow instability, 47.32.C- Vortex dynamics
% 47.54.De Experimental aspects


\maketitle

The wake flow behind a rigid obstacle is a central object of study
in fluid mechanics.  When the oncoming flow velocity exceeds a
threshold, vortices are shed behind the obstacle \cite{Tritton77}.
A typical wake is composed of successive eddies of alternating
sign -- the ``von K\'arm\'an vortex street" -- and is observed
over a wide range of flow velocities and body shapes
\cite{Dyke82,Williamson96}. The frequency of vortex shedding ($f$)
is determined by the flow velocity ($V$) and the object size
($d$), whose relation is captured by the near constancy of the
Strouhal number, $St=d\,f/V$ \cite{Williamson96}.

The dynamics of a rigid object which moves freely in the direction
perpendicular to the flow is of interest in many industrial and
biological applications \cite{Williamson04,Sarpkaya04,Liao03}.
Lateral motion of an object can be induced by interaction with the
flow and is often called the {\it vortex-induced vibration} (VIV).
At low flow velocities, the body starts to oscillate sideways with
small amplitude (less than 0.4 times body diameter). Its
associated wake structure is again a von K\'arm\'an vortex street.
However, further increase of flow velocity causes the obstacle to
oscillate in phase with the vortex shedding, and as a result, a
series of dipoles are shed instead \cite{Govardhan00,Brika93}.

Settling bodies or rising bubbles, where the balance of
gravitational and drag forces set the velocity, also exhibit
transitions as they interact with their wakes.  For example, a
slowly settling sedimenting sphere falls straight downwards
\cite{Nakamura76} but above a certain sedimentation velocity, the
sphere's motion becomes periodic and its trajectory a spiral or
zigzag \cite{Karamanev92}.  A {\it deformable} object, such as a
droplet or bubble, can behave similarly even as its shape now
changes \cite{Magnaudet00,Duineveld95}.

Finally, studies have shown the instability (and bistability) of
slender deformable bodies to lateral oscillations in quasi-2D
soap-film flows \cite{Zhang00}, and of heavy deformable sheets to
lateral oscillations in fast 3D flows
\cite{Shelley05,Watanabe02,Huang95,Taneda68}. In these cases, the
system corresponds to the flapping of a flag in a stiff breeze.

Flowing soap film provides a practical template upon which to
study the dynamics of a nearly 2D flow \cite{Couder89,Rutgers01}.
The experimental setup has been introduced earlier
\cite{Couder89,Zhang00,Alben02,Rutgers01}. In this work, we
introduce a deformable closed body into a fast flowing soap-film.
Two thin nylon wires (0.3 mm in diameter) separate at a nozzle
(0.5 mm inner diameter) attached to the bottom of a reservoir.
The reservoir contains soapy water maintained at a fixed pressure
head, thus fixing the flux.  A stopcock regulates the flow rate
through the nozzle. The nylon wires extend downwards to a
collection box 2.4 m below. Driven by gravity, the soap film flows
downwards. Owing to air drag, a terminal velocity is reached
approximately 60 cm below the nozzle with a velocity profile near
the center close to being uniform (velocity differences are within
20\,\% of the mean, over 60\,\% of the span about the midline).
From optical interference patterns, the film thickness is found to
vary smoothly across the film by about 15\,\% of its average
thickness.

%%%%%%%%%%%%%%%%%%%%%%%%%%%%%%%%%%%%%%%%%%%%%%%%%%%%%%%%%%%%
\begin{figure} %% fig_bistable.ai
    \centering
        \includegraphics[width=.5\textwidth]{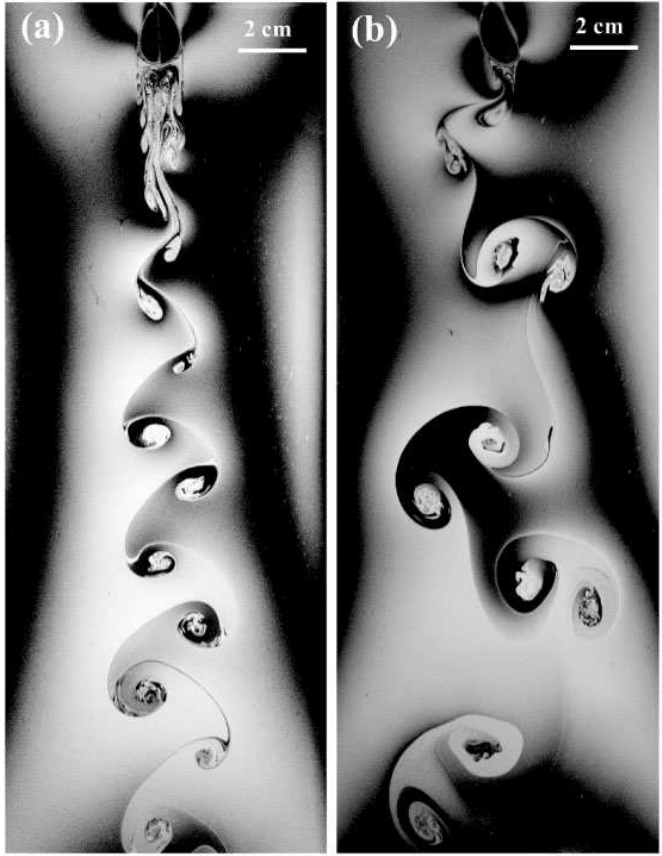}% {fig_bistable8.eps}
\caption{Flow structures behind a 5 cm loop at 2.2 m/s flow
velocity. The coupled fluid-structure system shows bistability:
(a) the stationary state; the loop remains essentially motionless
and its wake is a von K\'arm\'an vortex street. The loop is
deformed by the flow into a teardrop shape. (b) the oscillatory
state; the loop sheds two vortex dipoles within each oscillation
period. } \label{fig:bistable}
\end{figure}

We use a thin rubber loop (0.2 mm thick) as the deformable
structure. The loop wets into the soap-film and is supported from
its inner side against the flow. The loop is much thicker when
compared to the film thickness (0.003 mm) that the fluid
presumably does not penetrate over the loop. Six loops of
different circumferences (5--7.5 cm) are used. We find that for
regimes studied here, the loop appears to undergo only bending
deformations, and not stretching or compression, as its length
shows no measurable increase or decrease. Currently, we do not
understand what balance of effects sets the enclosed area of the
loop, which is an important constraint on the possible dynamics.
However, we do find that, once experimental conditions are fixed,
and the loop is in a fixed state of dynamics, the enclosed area
changes very little in time (e.g. $\sim5\,\%$ for a 5 cm loop).
However, between different states or conditions, the enclosed area
can vary by factors of two or three.

A laser Doppler velocimeter (LDV; Model LDP-100, TSI Inc.) is used
to record the upstream velocity $V$. Micron-sized particles
(TiO$_2$) are seeded into the flow for LDV measurements. Flow
structures are visualized using interference patterns from
monochromatic illumination (low-pressure sodium lamps operating at
wavelength 585 nm). The movies of the wake flow together with the
loop are recorded using a high speed camera at 1000 frames per
second.

The interaction between the loop and the flow is quite
complicated. In our experiments, we observe bistable states, one
stationary and another oscillatory (see Fig.\,\ref{fig:bistable}a
and b), that co-exist over a range of flow velocities. At least in
the conditions considered here, we do not observe spontaneous
transitions between these two states.  However, a transition from
the stationary to the oscillatory state can be induced by
externally perturbing the loop, or by abruptly changing the flow
velocity.

%%%%%%%%%%%%%%%%%%%%%%%%%%%%%%%%%%%%%%%%%%%%%%%%%%%%%%%%%%%%%%%%%%
\begin{figure} %% loop_and_centroid.m
  \centering
  \includegraphics[width=.5\textwidth]{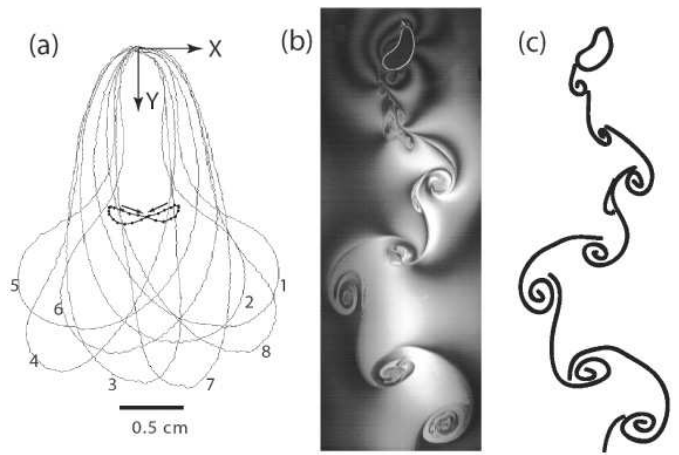}% {Fig_centroid_XY_fig_long.eps}% {Fig_centroid_XY_fig2.eps} % {Fig_centroid_XY.eps}
\caption{ (a) At 2.2 m/s flow velocity, eight snapshots of a 5 cm
loop and its centroid are shown. The oscillation period of the
oscillating loop is 52 ms. The snapshots of the flexible loop are
presented as solid loops at 6 ms intervals and the trajectory of
the centroid of the loop with 2 ms intervals. The loops are
sequentially numbered. (b) Wake structure when the loop centroid
is at the far left. The loop starts to shed a counter-clockwise
vortex. The wake structure is schematically shown in (c).}
   \label{fig:centroid}
\end{figure}

In the stationary state, the loop behaves as a rigid hoop and has
a teardrop shape with higher curvature on the top than on the
sides (Fig.\,\ref{fig:bistable}a).  A characteristic length scale
($D$) of the loop, its width in the film, is 1 cm.  The flow
velocity ($V$) varies from 1.5 to 2.5 m/s and the kinematic
viscosity ($\nu$) of soap film is approximately 0.04 cm$^2$/s. The
frequency of vortex shedding ($f\sub{s}$) varies from 20 to 50 Hz.
Based on these characteristic numbers, we estimate the Reynolds
number ($Re$) and Strouhal number ($St_s$) for the system to be
\beq
Re = \f{V D}{\nu} \sim 5,000 \,, ~~ St_s = \f{ D
f\sub{s}}{V} \sim 0.2 \,,
\eeq
where the subscript $s$ stands for vortex shedding, since the
Strouhal number is calculated based on the vortical structure of
the wake. Figure\,\ref{fig:bistable}a shows the deforming body and
its vortical wake using a 5 cm loop and flow velocity of 2.2 m/s.
As can be seen, vortices of alternating sign are successively
produced.

In the co-existing oscillatory state, shown in
Fig.~\ref{fig:bistable}b at the same parameters as above, the loop
now oscillates periodically in the horizontal direction, and the
vortical wake behind it is quite different.  For low flow
velocity, two dipole pairs are shed during each oscillation
period. Such a wake structure is also observed behind oscillating
cylinders and is referred to as the {\it 2P mode} \cite{Brika93}.

%%%%%%%%%%%%%%%%%%%%%%%%%%%%%%%%%%%%%%%%%%%%%%%%%%%%%
\begin{figure} %% check_movie.m
    \centering
        \includegraphics[width=.5\textwidth]{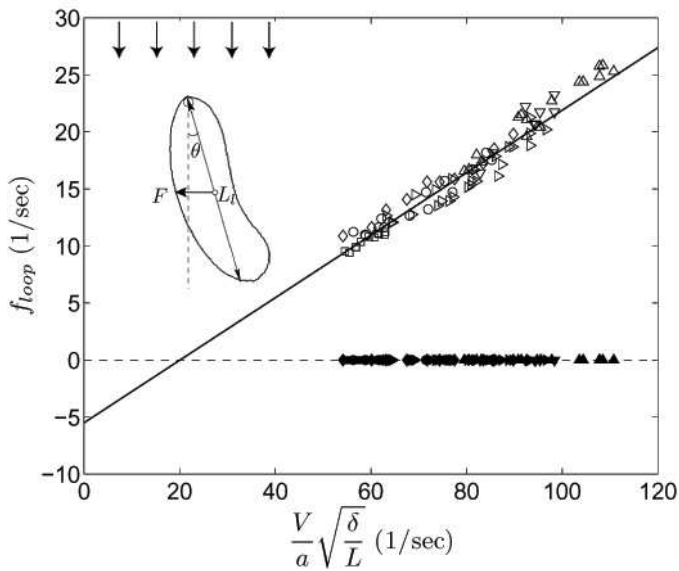}% {Fig_linear2.eps}
\caption{Frequency of an oscillating loop ($f_{loop}$) versus
rescaled velocity ($V \sqrt{\delta}/a \sqrt{L}$). We test over 6
different loop lengths (5 cm ($\bigtriangleup$), 5.5 cm
($\bigtriangledown$), 6 cm ($\rhd$), 6.5 cm ($\Diamond$), 7 cm
($\square$), and 7.5 cm ($\bigcirc$)). The open symbols are in the
oscillatory state and the closed ones are in the stationary state.
The frequency of the oscillating loop is linearly proportional to
the rescaled velocity.} \label{fig:linear}
\end{figure}
%%%%%%%%%%%%%%%%%%%%%%%%%%%%%%%%%%%%%%%%%%%%%%%%%%%%%%%%%%5

Figure\,\ref{fig:centroid}a shows both the position of the loop at
several time-points during one period of oscillation, and the path
taken by its centroid.  During the oscillation, the loop
continuously changes its shape, and its centroid moves along a
figure-eight trajectory (Fig.\,\ref{fig:centroid}a). This
figure-eight shape is due to the fact that the frequency of
oscillation in the stream-wise direction is twice that in the
transverse direction.  This has been observed in the motions of a
flapping flag \cite{Zhang00} and VIV systems
\cite{Sarpkaya04,Kim02}. Also, the loop oscillates in phase with
that vortex shedding (Fig.\,\ref{fig:centroid}b and c). When the
loop is at far right (or left), a clockwise turning (or
counter-clockwise) vortex is shed.

By using loops of several different lengths, we find a linear
relation between the oscillation frequency of the loop
($f_{loop}$) and a rescaled velocity $V\sqrt{\delta} / a\sqrt{L}$
(see Fig.\,\ref{fig:linear}), where $\delta$ is the film
thickness, $a$ the thickness of the loop, and $L$ the loop length.
Our results from loops of different lengths and differing flow
velocities all collapse onto a single line with slope of about
0.27.  This offset of this affine relation suggests a bifurcation
to oscillation at a finite flow velocity;  linear extrapolation to
$f_{loop} = 0$ suggests a critical rescaled velocity of about 20,
which is unfortunately beyond the reach of this experiment.

To better understand the relation between oscillation frequency
and flow velocity, we propose a simple model for the oscillations
of an elongated loop with longitudinal length $L_l$ driven by a
``lift force'' in the direction perpendicular to the stream. The
lift force ($F$) is taken as proportional to $\rho V^2 L_l \delta$
where $\rho$ is the density of fluid, $V$ the fluid velocity.
Hence, $F = ({1}/{2}) C_L \rho V^2 L_l \delta $, where $C_L$ is a
lift coefficient. Typically, $L_l$ is proportional to the loop
circumference, $L$. At an angle $\theta$ inclined to the flow
stream, $C_L$ is proportional to $\sin \theta$ \cite{Landau59}.
For small $\theta$, $\sin \theta \sim {x\sub{cm}}/{y\sub{cm}}$
where $(x\sub{cm},y\sub{cm}$ is the center of mass (centroid)
location.  Therefore, we approximate the lift force as $F = m
\ddot{x}\sub{cm} \approx (1/2)\rho V^2 L_l \delta
{x\sub{cm}}/{y\sub{cm}} $, where $\ddot{x}\sub{cm}$ is the
acceleration in the transverse direction and $m$ is the total body
mass. In this experiment, the mass of the (wetted) loop is much
greater than that of the enclosed fluid.  Hence, we assume that
the total body mass ($m$) is proportional to $\rho_L L a^2$ where
$\rho_L$ is the density of the loop.  Also, the $y$-component of
the centroid, $y\sub{cm}$, and the length $L_l$ are assumed to be
proportional to the length of the loop if the body is elongated
due to the flow. With the trivial solution for the $x$-component
of centroid as $x\sub{cm}= C e^{i \omega t}$, we obtain an
expression for the oscillation frequency: $\omega = 2 \pi
f\sub{loop} \propto V\sqrt{\delta}/a \sqrt{L}$.  This is
consistent with our observations and underlies our rescaling of
the data in Fig.~\ref{fig:linear}. Put differently, this is simply
the oscillation frequency of a hanging pendulum where the
gravitational force is replaced by a drag force.

%%%%%%%%%%%%%%%%%%%%%%%%%%%%%%%%%%%%%%%%%%%%%%%%%%%%%
\begin{figure} %% check_movie.m
    \centering
        \includegraphics[width=.5\textwidth]{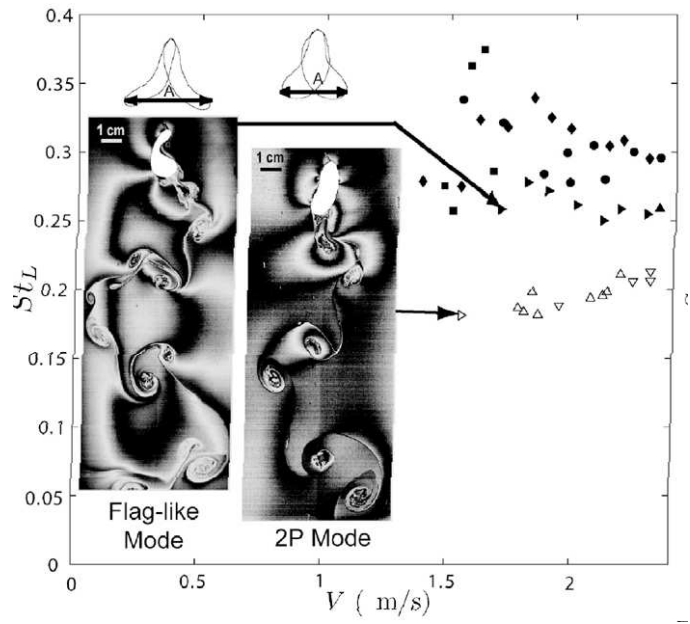}%{Fig_strouhal_wA3.eps}
\caption{Strouhal number of the loop ($St_L$) versus the flow
velocity ($V$) for different loop lengths (5 cm
($\bigtriangleup$), 5.5 cm ($\bigtriangledown$), 6 cm ($\rhd$),
6.5 cm ($\Diamond$), 7 cm ($\square$), and 7.5 cm ($\bigcirc$)).
Open symbols are 2P mode and closed ones are flag-like modes. The
Strouhal numbers of the two modes are well separated, and the
Strouhal numbers of 2P modes are close to 0.2 whereas those of
flag-like modes are above 0.25.} \label{fig:strouhal}
\end{figure}

As the flow velocity increases, a more complicated mode in the
oscillatory state can be observed (left panel in
Fig.\,\ref{fig:strouhal}). In this case, the loop sheds more than
four vortices over a single period of oscillation. We refer to
this wake structure as a {\it flag-like mode}.  We use this
terminology because the wake now resembles more that behind a
flapping flag (see \cite{Zhang00}), and because the body itself
looks elongated and ''flag-like''.  This is because the enclosed
area is now smaller in relation to $L^2$ than for the body in the
2P mode. To characterize the loop oscillation and the wake
structure, we again define a Strouhal number, now using the
oscillation frequency of the loop itself, or $St_L =
Af\sub{loop}/V$ where $A$ is the outer amplitude of oscillation as
indicated in Fig.\,\ref{fig:strouhal}. The corresponding Strouhal
numbers for 2P modes and for flag-like modes are shown in
Fig.\,\ref{fig:strouhal}.  The 2P mode yields $St_L$ approximately
0.2 and the flag-like modes yield values above 0.25.

\begin{figure} %%
    \centering
        \includegraphics[width=.5\textwidth]{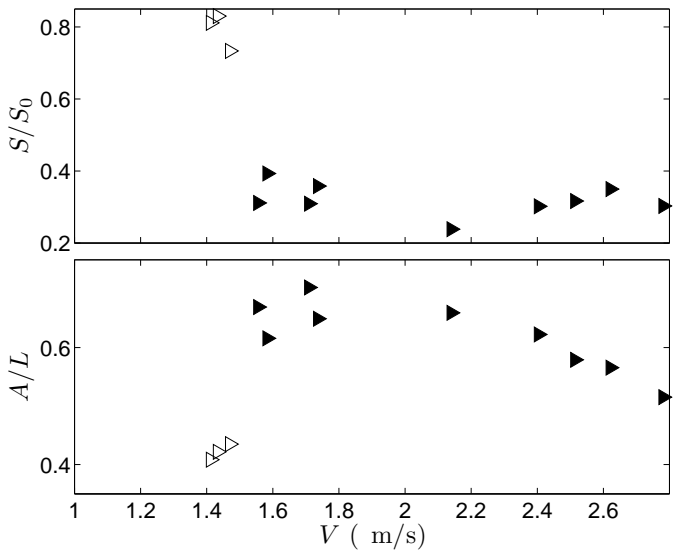}%{Paper_Ampl.eps}
\caption{Enclosed area (normalized by $L^2/4\pi$) and amplitude
(normalized by $L$) of the oscillating loop for 6 cm loop. Area
and amplitude abruptly change as the wake structure transits from
2P modes (open symbols) to flag-like modes (closed symbols). As
the velocity increases further, the amplitude decreases due to the
large drag force.}
    \label{fig:ampl}
\end{figure}

This separation in the respective Strouhal numbers is caused by a
discontinuous change of the amplitude $A$ from the 2P mode to the
flag-like mode. Figure\,\ref{fig:ampl} shows the simultaneous
transitions of enclosed area ($S$) and amplitude ($A$) of the
oscillating loop. At high flow velocity, the streamwise extension
of the loop increases due to the higher drag, and this presumably
causes the observed relative decrease in enclosed area in the
flag-like mode. Due to its now higher aspect ratio, the loop's
amplitude increases (compare left and right panels in
Fig.\,\ref{fig:strouhal}).  The causes of this abrupt change in
enclosed loop area and oscillation amplitude remains an open
question. Following this abrupt change, the frequency of vortex
shedding also increases presumably due to the higher aspect ratio.
Unlike the 2P mode, the oscillation of the loop and vortex
shedding are not in-phase, and the wake becomes more complicated.

We have reported on the dynamics of a flexible body as it
interacts with an impinging high-speed flow. We find that the loop
can have co-existing stable states. In the oscillatory state, the
loop oscillation frequency is linearly proportional to the flow
velocity and inversely proportional to the square root of the loop
length. Also, it has been observed in a system with a freely
vibrating solid cylinder that oscillation frequency linearly
depends on the upstream flow velocity \cite{Govardhan00,Brika93}.
This relationship can be explained by a simple model based on the
lift force of an inclined airfoil. With small loop lengths and low
flow velocity, a 2P wake mode (with pairs of shed vortices
\cite{Govardhan00,Brika93}) is observed and has constant Strouhal
number $\sim 0.2$. At longer loop lengths and higher flow
velocity, a flag-like mode appears with higher Strouhal number
($\ge$ 0.25). This separation in Strouhal numbers is a result of
abrupt changes in enclosed loop area and oscillation amplitude.

This work is supported by DOE Grant No. DE-FG02-88ER25053.

\end{document}